\documentclass[a4paper]{article}
\usepackage{pgfplots}
\usetikzlibrary{patterns}
\usepackage{INTERSPEECH2018}
\usepackage{amsmath,graphicx}
\usepackage{amsfonts}

\title{Investigation of using disentangled and interpretable representations \\
for one-shot cross-lingual voice conversion}
\name{Seyed Hamidreza Mohammadi, Taehwan Kim}
\address{
  ObEN, Inc.}
\email{hamid@oben.com, taehwan@oben.com}

\begin{document}

\maketitle
\begin{abstract}
We study the problem of cross-lingual voice conversion in non-parallel speech corpora and one-shot learning setting. Most prior work require either parallel speech corpora or enough amount of training data from a target speaker. However, we convert an arbitrary sentences of an arbitrary source speaker to target speaker's given only one target speaker training utterance. To achieve this, we formulate the problem as learning disentangled speaker-specific and context-specific representations and follow the idea of \cite{hsu2017unsupervised} which uses Factorized Hierarchical Variational Autoencoder (FHVAE). After training FHVAE on multi-speaker training data, given arbitrary source and target speakers' utterance, we estimate those latent representations and then reconstruct the desired utterance of converted voice to that of target speaker. We investigate the effectiveness of the approach by conducting voice conversion experiments with varying size of training utterances and it was able to achieve reasonable performance with even just one training utterance. We also examine the speech representation and show that World vocoder outperforms Short-time Fourier Transform (STFT) used in \cite{hsu2017unsupervised}. Finally, in the subjective tests, for one language and cross-lingual voice conversion, our approach achieved significantly better or comparable results compared to VAE-STFT and GMM baselines in speech quality and similarity.   
\end{abstract}

\vspace{-2.5mm}
\section{Introduction}
\vspace{-1.5mm}
The task of Voice Conversion (VC) \cite{stylianou2009voice,mohammadi2017overview} is a technique to convert source speaker's spoken sentences into those of a target speaker's voice. It requires to preserve not only the target speaker's identity, but also phonetic context spoken by the source speaker. To tackle this problem, many approaches have been proposed \cite{toda2007voice, desai2010spectral, wu2012mixture}. However, most prior work require parallel spoken corpus and enough amount of data to learn the target speaker's voice. Recently, there were approaches proposed for voice conversion with non-parallel corpus \cite{hsu2016voice, song2013non, hsu2017voice}. But they still require that speaker identity was known \emph{priori}, or included in training data for the model. 

Recently, Hsu et al. \cite{hsu2017unsupervised} proposed to use disentangled and interpretable representations to overcome these limitations by exploiting Factorized Hierarchical Variation Autoencoder. They achieved reasonable quality with just single utterance from a target speaker but it was still not satisfactory. Nevertheless, most prior work focus on voice conversion \emph{within} one language. But we believe that if we can capture disentangled representations of phonetic or linguistic contexts and speaker identities, the model should be capable for more challenging \emph{cross}-lingual setting, which means that source and target speakers are from different languages. Therefore, we focus on investigating cross-lingual voice conversion, and propose to follow the same spirit from Hsu et al. \cite{hsu2017unsupervised} and improve the performance. Our contributions are:
\begin{itemize}
\item We investigate the different feature representations for spoken utterances by considering  Mel-cepstrum (MCEP) features and other acoustic features, and achieve better results compared to baselines.
\item We examine the effect of the size of training utterances from source and target speakers, and demonstrate that with just a few, or even one, utterances, we are able to achieve the reasonable performance.  
\item We conduct cross-lingual voice conversion experiments and our approach achieved significantly better or comparable results than baselines in speech quality and similarity in the subjective tests.
\end{itemize}

\vspace{-4mm}
\section{Related Work}
\vspace{-1mm}
Voice conversion has been an important research problem for over a decade. One popular approach to tackle the problem is spectral conversion such as Gaussian mixture models (GMMs) \cite{toda2007voice} and deep neural networks (DNN) \cite{desai2010spectral}. However, it requires parallel spoken corpus and dynamic time warping (DTW) is usually used to align source and target utterances. To overcome this limitation, non-parallel voice conversion approaches were proposed, for instance, \emph{eigenvoice} \cite{wu2012mixture}, \emph{i-vecotor} \cite{kinnunen2017non}, and Variational Autoencoder \cite{hsu2016voice, hsu2017voice} based models. However, eigenvoice based approach \cite{wu2012mixture} still requires reference speaker to train the model, and VAE based approaches \cite{hsu2016voice, hsu2017voice} require speaker identities to be known priori as included in training data for the model. i-vector based approach \cite{kinnunen2017non} looks promising which remains to be studied further. The i-vectors are converted by replacing the source latent variable by the target latent variable. The Gaussian mixture means are then reconstructed from the converted i-vector. The Gaussians with adjusted means are then applied to the source vector to perform the acoustic feature conversion. Siamese autoencoder has also been proposed for decomposing speaker identity and linguistic embeddings \cite{mohammadi2017siamese}. However, this approach requires parallel training data to learn the decomposing architecture. This decomposition is achieved by means of applying some similarity and non-similarity costs between the Siamese architectures.

Nonetheless, cross-lingual voice conversion is also a challenging task since target language is not known in training time, and only few work has proposed, including GMMs based approach \cite{ramani2016multi} and eigenvoice based approach \cite{charlier2009cross}, but still have inherent limitations as above.

Recently, deep generative models have been applied and successful for unsupervised learning tasks, and include Variational Autoencoder (VAE) \cite{kingma2013auto}, Generative Adversarial Networks (GAN) \cite{goodfellow2014generative}, and auto-regressive models \cite{oord2016pixel, van2016wavenet}. Among them, VAE can infer latent codes from data and generate data from them by jointly learning inference and generative networks, and VAE has been also applied for voice conversion \cite{hsu2016voice, hsu2017voice}. However, in their models, speaker identities are not infered from data and instead required to be known in model training time. GAN has been also exploited for non-parallel voice conversion \cite{kaneko2017parallel} with the cycle consistency contraint \cite{zhu2017unpaired}, but it still has the limitation that it needs to know the target speaker in training time and be trained for each target.

To understand the disentangled and interpretable structure of latent codes, several work were proposed, namely, DC-IGN \cite{kulkarni2015deep}, InfoGAN \cite{chen2016infogan}, $\beta$-VAE \cite{higgins2016beta}, and FHVAE \cite{hsu2017unsupervised}.  These approaches to uncover disentangled representation may help voice conversion with very limited resource from target speaker, since it might infer speaker identity information from data without supervision, as illustrated in FHVAE \cite{hsu2017unsupervised}. However, the qualities of converted voices were not good enough, therefore, we focus on the model structure of FHVAE and investigate to improve it, even with more challenging cross-lingual voice conversion setting.


\vspace{-3mm}
\section{Model}
\vspace{-1.5mm}


Variational autoencoder \cite{kingma2013auto} (VAE) is a powerful model to uncover hidden representation and generate new data samples. Let observations be $x$ and latent variables $z$. In the variational autoencoder model, the encoder (or inference network) $q_\phi (z|x)$ outputs $z$ given input $x$, and decoder $p_\Phi (x|z)$ generates data $x$ given $z$. The encoder and decoder are neural networks. Training is done by maximizing variational lower bound (or also called evidence lower bound): 
\vspace{-2mm}
\begin{align*}    
    \ell (\Phi, \phi) & = \mathbb{E}_q[\log p_\Phi(x,z)] - \mathbb{E}_q [\log q_\phi (z|x)] \\
    &= \log p_\Phi(x) - D_{KL}( q_\phi (z|x) || p_\Phi(z|x)).
\end{align*}
where $D_{KL}$ is Kullback-Leibler divergence.

\begin{figure}
\begin{center}
\vspace{-5mm}
\includegraphics[width=0.3\textwidth]{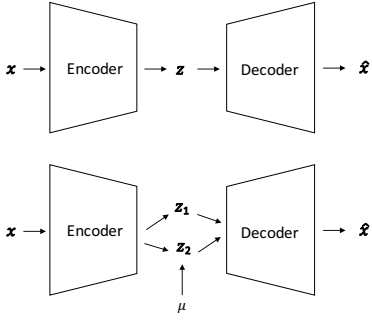} 
\vspace{-8mm}
\caption{Structures of Variation Autoencoder (upper) and Factorized Hierarchical Variational Autoencoder (lower).}
\vspace{-5mm}
\label{f:model_vae}
\end{center}
\end{figure}
However, VAE considers no structure for latent variable $z$. Assuming structure for $z$ could be beneficial to exploit the inherent structures in data. Here we describe Factorized Hierarchical Variational Autoencoder proposed by Hsu et al \cite{hsu2017unsupervised}. Let a dataset $D$ consist of $N_{seq}$ i.i.d. sequences $X^i$. For each sequence $X^i$, it consists of $N_{seg}^i$ $X^{i,j}$ observation segments. 
Then we define factorized latent variables of latent segment variable $Z^{i,j}_1$ and latent sequence variable $Z^{i,j}_2$. 
In the context of voice conversion, $Z^{i,j}_1$ is responsible for generating phonetic contexts and $Z^{i,j}_2$ is for speaker identity. When generating data $X^{i,j}$, we first sample $Z^{i,j}_2$ from isotropic Gaussian centered at $\mu^i$ shared for the entire sequence, and also $Z^{i,j}_1$ independently. Then we generate $X^{i,j}$ conditioned on $Z^{i,j}_1$ and $Z^{i,j}_2$. Thus, joint probability with a sequence $X^i$ is:
\vspace{-2mm}

\begin{align*}
p_\Phi (X^i, Z_1^i, Z_2^i, \mu^i) = p_\Phi (\mu^i) \prod_{j=1}^{N_{seg}^i} & p_\Phi(X^{i,j}|Z_1^{i,j}, Z_2^{i,j}) \\
&p_\Phi (Z_1^{i,j}) p_\Phi (Z_2^{i,j}|\mu^i)
\vspace{-2mm}
\end{align*}
This is illustrated in Figure \ref{f:model_vae}. For inference, we use variational inference to approximate the true posterior and have:
\vspace{-2mm}
\begin{align*}
q_\phi ( Z_1^i, Z_2^i, \mu^i | X^i) = q_\phi (\mu^i) \prod_{j=1}^{N_{seg}^i} & q_\phi(Z_1^{i,j}|X^{i,j}, Z_2^{i,j}) \\
& q_\phi(Z_2^{i,j}|X^{i,j})  
\vspace{-2mm}
\end{align*}
Since sequence variational lower bound can be decomposed to segment variational lower bound, we can use batches of segment instead of sequence level to maximize:
\vspace{-2mm}
\begin{align*}
\ell(\Phi, \phi; X^{i,j}) = \ell(\Phi, \phi; X^{i,j} | \tilde{\mu}^i) + \frac{1}{N_{seg}^i}\log p_\Phi (\tilde{\mu}^i) + const &\\
\ell(\Phi, \phi; X^{i,j} | \tilde{\mu}^i) =  \mathbb{E}_{q_\phi(Z^{i,j}_1,Z^{i,j}_2|X^{i,j})}[\log p_\Phi (X^{i,j} | Z_1^{i,j}, Z_2^{i,j})] &\\
                                                           - \mathbb{E}_{q_\phi(Z^{i,j}_2|X^{i,j})} [D_{KL}(q_\phi(Z_1^{i,j}|X^{i,j}, Z_2^{i,j}) || p_\Phi(Z_1^{i,j}))] &\\
                                                           - D_{KL}(q_\phi(Z_2^{i,j}|X^{i,j}) || p_\Phi(Z_2^{i,j}|\tilde{\mu}^i))&
\vspace{-2mm}
\end{align*}
where $\tilde{\mu}^i$ is the posterior mean of $\mu^i$. Please refer to Hsu et al. \cite{hsu2017unsupervised} for more details. Additionally, Hsu et al. also proposed discriminative segment variational lower bound to encourage $Z_2^i$ to be more sequence-specific by adding the additional term of inferring the sequence index $i$ from $Z_2^{i,j}$. For our experiments, we exploit this FHVAE model and sequence-to-sequence model \cite{sutskever2014sequence} as the structure of encoder-decoder for sequential data. 

For performing the voice conversion, we compute the average $Z_2$ from the training utterance(s) of source and target speakers. For a given input utterance, we compute $Z_1$ and $Z_2$ of the input utterance. There are two ways to perform voice conversion. First, we can replace $Z_2$ values of the source speaker with the average $Z_2$ from the target speaker. This approach resulted in too muffled generated result.  Second, we  compute a difference vector between source and target average $Z^{diff}_2=Z^{trg}_2-Z^{src}_2$. This difference vector is added to $Z_2$ from the input utterance as $Z^{converted}_2=Z_2+Z^{diff}_2$ and then decoded using FHVAE to achieve the speech features. In an informal listening test, we decided to the second approach since it resulted in significantly higher quality generated speech.

\vspace{-1.5mm}
\section{Experiments}
\vspace{-1.5mm}
\subsection{Datasets}
\vspace{-1.5mm}
We used the TIMIT corpus \cite{garofolo1993darpa} which is a multi-speaker speech corpus as the training data for FHVAE model. We used the training speakers as suggested by the corpus to train the model. For English test speakers, we select speakers from TIMIT testing part of the corpus. We also use a proprietary Chinese speech corpus (hereon referred to as CH) with 5200 speakers each uttering one sentence. We consider using the combination of TIMIT and Chinese corpus for training the model as well. For Chinese test speakers, we utilize speakers from the THCHS-30 speech corpus \cite{wang2015thchs}. To observe the effect of having more utterances per speaker but less speakers we also train the model on VCTK corpus~\cite{veaux2017cstr}. Finally, for objective testing (which requires availability of parallel data), we utilized four CMU-arctic voices (BDL, SLT, RMS, CLB)\cite{kominek2004cmu}. As speech features, we used 40th-order MCEPs (excluding the energy coefficient, dimensionality D=39), extracted using the World toolkit \cite{morise2016world} with a 5ms frame shift. All audio files are transformed to 16kHz and 16 bit before any analysis. 
\vspace{-3mm}
\subsection{Experimental setting}
\vspace{-1.5mm}
For the encoder and decoder in FHVAE model, we use Long Short Term Memory (LSTM) \cite{hochreiter1997long} as the first  layer with 256 hidden units with a fully-connected layer on top. We use 32 dimensions for each latent variable $Z_1$ and $Z_2$. The models were trained with stochastic gradient descent. We use a mini-batch size of 256. The Adam optimizer \cite{kingma2014adam} is used with $\beta_1$ = 0.95, $\beta_2$ = 0.999, $\epsilon$ = $10^{-8}$, and initial learning rate of $10^{-4}$. The model is trained for 500 epochs and select the model best performing on the development set.

From now on, we use the abbreviation VAE for FHVAE model. In our experiments, we consider three models: GMM (GMM MAP Adaptation \cite{toda2007voice}), VAE-STFT (uses STFT as speech analysis/synthesis\cite{hsu2017unsupervised}), VAE (uses World as speech analysis/synthesis\cite{morise2016world}). We consider four gender conversions (F: female, M: male): F2F, F2M, M2F, M2M. We also consider four cross-language conversions (E: English, Z: Chinese): E2E, E2Z, Z2E, Z2Z. The voice conversion samples are available at: https://shamidreza.github.io/is18samples
\vspace{-3mm}
\subsection{Visualizing embeddings}
\label{vis_emb}
\begin{figure}[t]
\begin{center}
\includegraphics[width=0.30\textwidth]{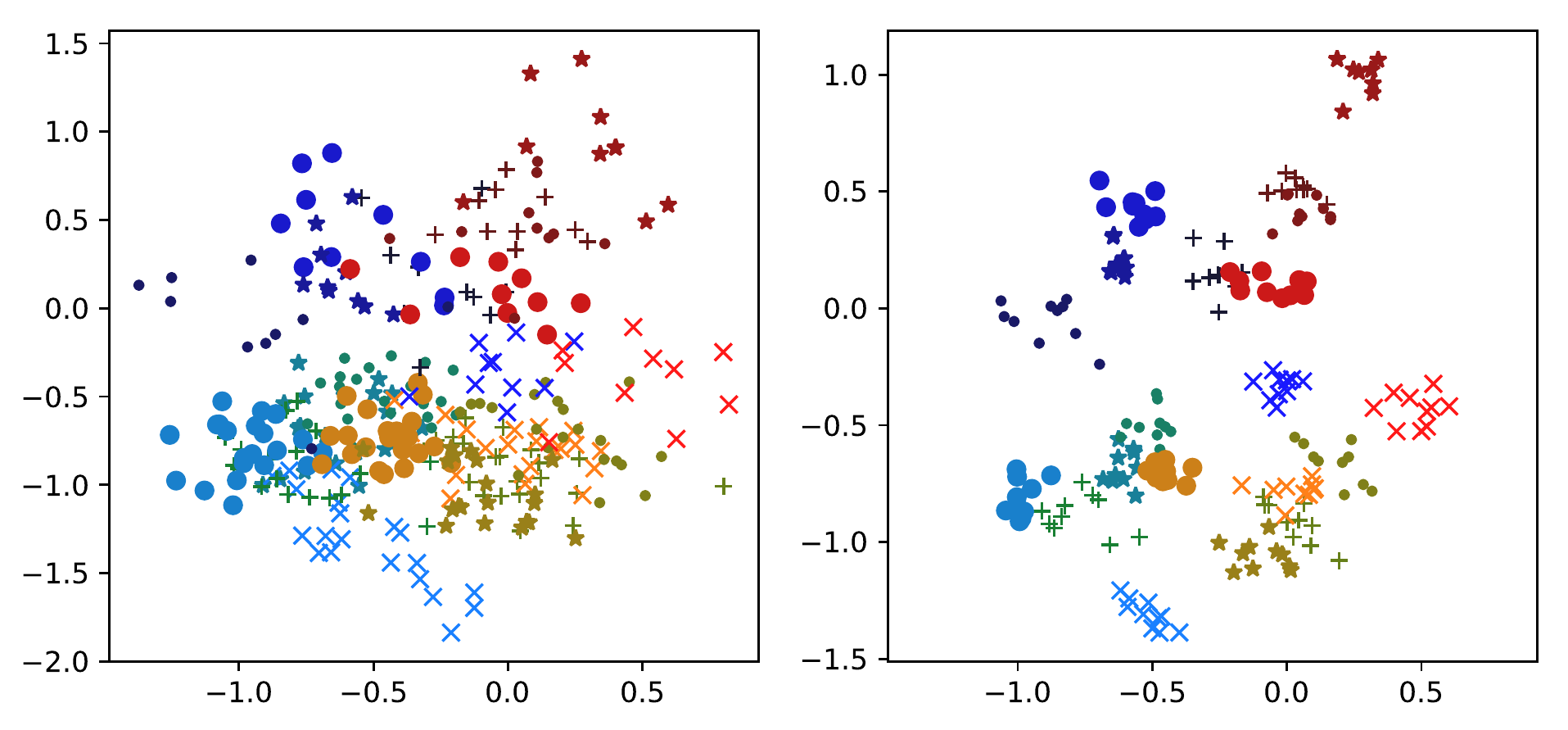} 
\includegraphics[width=0.30\textwidth]{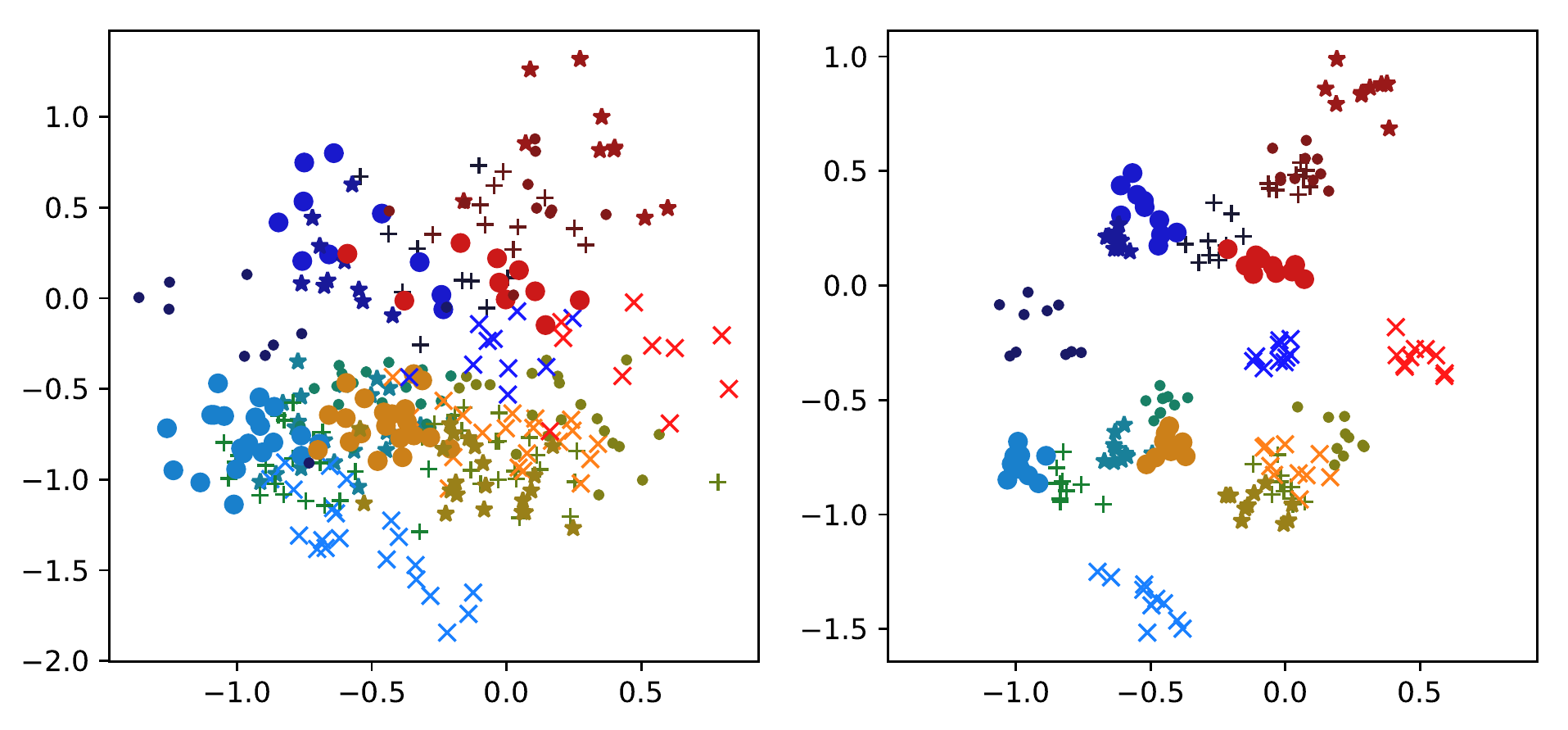} 
\includegraphics[width=0.30\textwidth]{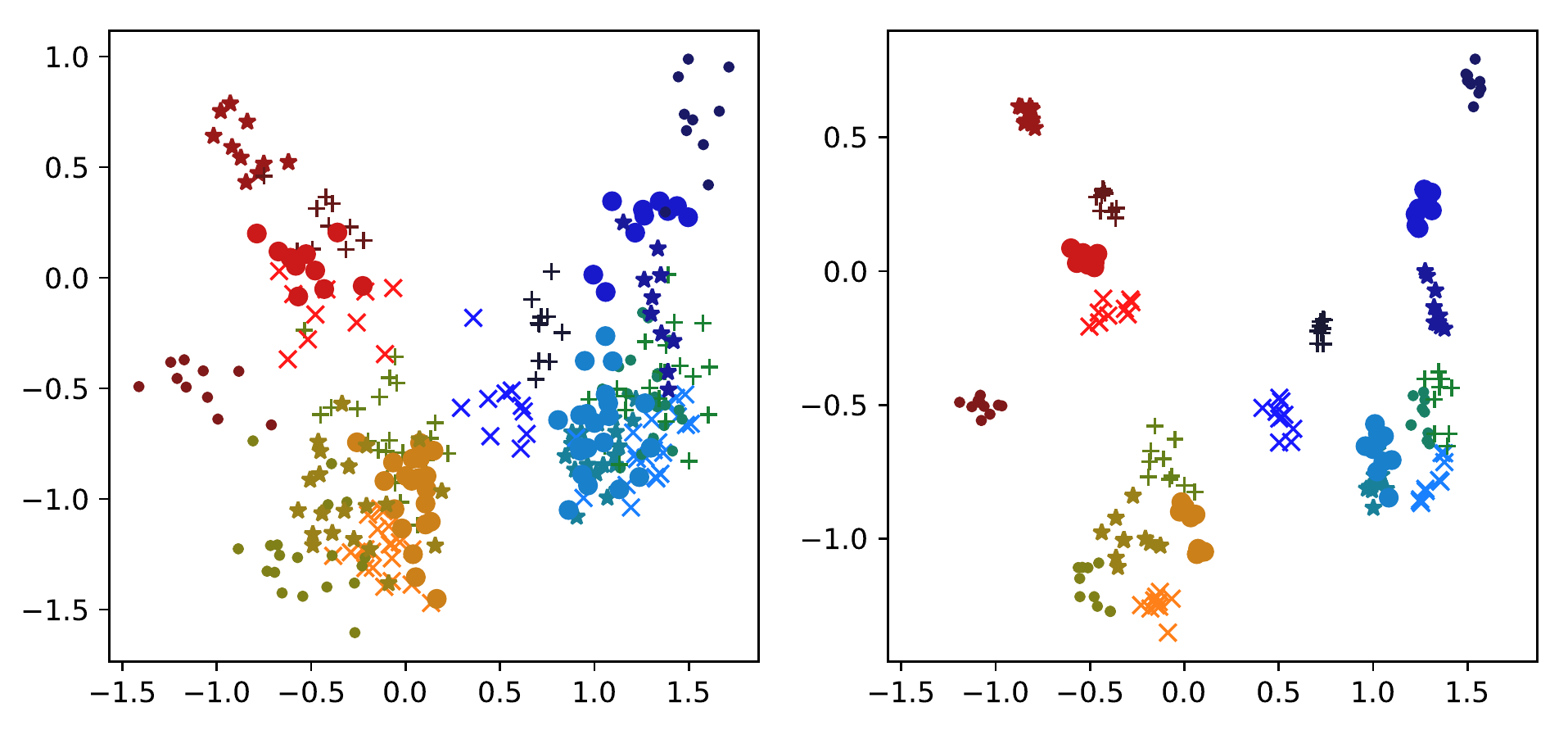} 
\caption{Visualization of speaker embedding. Each point represents single utterance and different colors represent different speaker/languages; blueish dots are English females and light blueish are Chinese females; and reddish dots are English males and orange dots are Chinese males. See Section \ref{vis_emb} for details.}
\vspace{-2.5mm}
\label{f:spk}
\end{center}
\vspace{-5mm}
\end{figure}

\begin{figure}[t]
\vspace{-0.5mm}
\begin{center}
\includegraphics[width=0.32\textwidth]{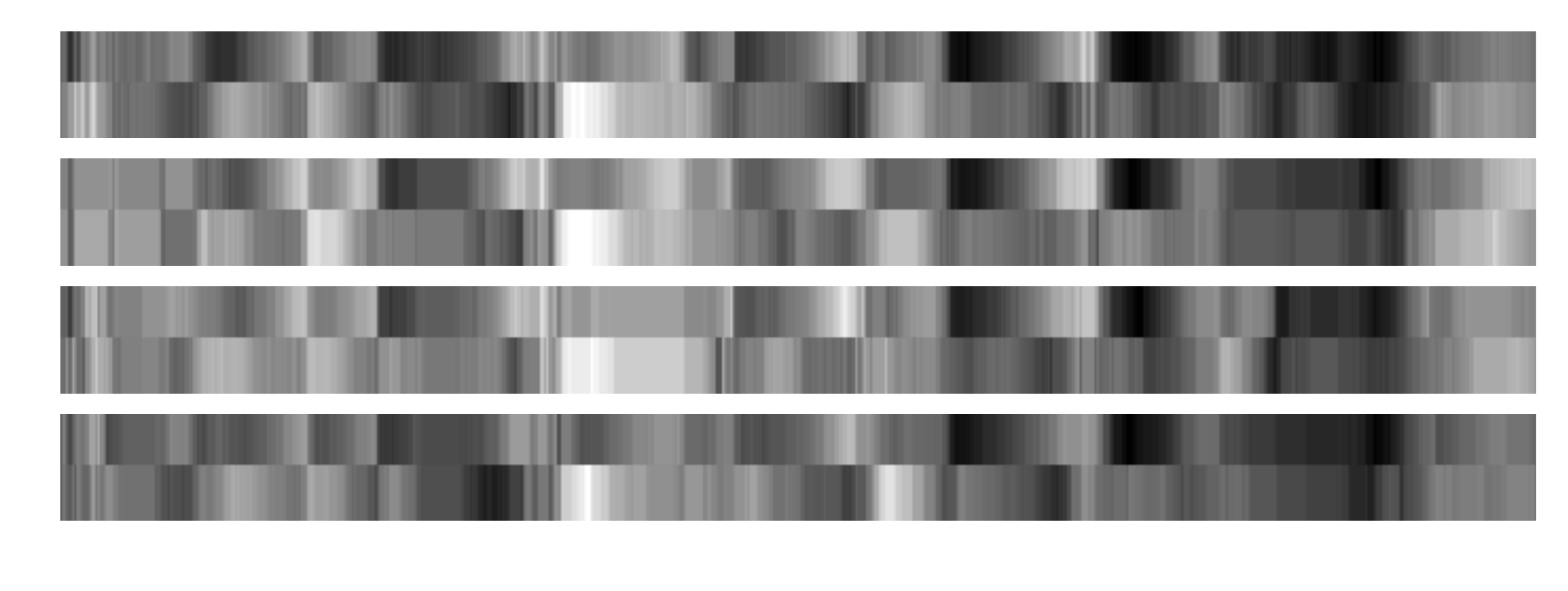} 
\vspace{-5mm}
\caption{Visualization of phonetic context embedding sequence of the sentence "She had your dark suit in greasy wash water all year" aligned to each other for two female speakers (top) and two female speakers (bottom). The embeddings are transformed to 2D using PCA.}
\label{f:phon}
\end{center}
\vspace{-11.5mm}
\end{figure}

\vspace{-1.5mm}
In this experiment, we investigate the speaker embeddings $Z_2$ by visualizing them in Figure \ref{f:spk}. For visualizing the speaker embeddings, we use the 10 test speakers from TIMIT test set (red data points for males and blue for female) and 10 test speakers from THCHS-30 (orange/greenish data points for males and light blue for female). We also use VAE models trained on TIMIT (top), CH corpus (middle), and TIMIT+CH corpus (bottom). In Figures \ref{f:spk}, we show the speaker embedding from 1 sentence in the left subplots and from 5 sentences in the right subplots, where the 2D plot of the speaker embeddings (computed using PCA) are shown. In all subplots, the female and male embedding cluster locations are clearly separated. Furthermore, the plot shows that the speaker embeddings of unique speakers fall near the same location. Although when 5 utterances are used to compute the embedding value, the variation is visibly less compared to when merely one sentence is used. This shows sensitiveness of the speaker embedding computed from the model to sentence variations. Also it is interesting to note that when both TIMIT+CH corpus are used for training, the speaker embeddings are further apart suggesting a better model property. One phenomenon that we notice is that the speaker embeddings for different languages and gender fall to different locations. This shows the embeddings are still language dependent, which might suggest the network learn to use the phonetic information to learn some language embedding within $Z_2$ as an additional factor. Furthermore, we investigate the phonetic context embedding $Z_1$ for a sentence for four test speakers on TIMIT-trained VAE. The phonetic context matrix over the computed utterances (compressed using PCA) is shown in Figures \ref{f:phon}. Ideally, we want the matrices should be close to each other since the phonetic context embedding is supposed to be speaker-independent. The figure show the closeness of the embeddings at the similar time frames. There is still some minor discrepancy between the embeddings which shows room for further improvement of model architecture and/or larger speech corpus.

\vspace{-4.5mm}
\subsection{Effect of training data size}
\begin{figure}[t]

\centering

\begin{tikzpicture} 
	\begin{axis}[
		height=3cm,
		width=8cm,
		xmin=1,
		xmax=100,
		ylabel={mel-CD (dB)},
		ylabel near ticks,
		legend style={font=\scriptsize,at={(0.5,0.95)},anchor=north},
		grid=major,
	] 	
	\addplot[color=black,dotted, ultra thick] coordinates {
		(1,8.66)
		(2,8.52)
		(5,8.25)
		(10,7.99)
		(20,7.87)
		(50,7.75)
		(100,7.69)		
	};
	\addlegendentry{GMM conversion}

	\addplot[color=black,solid] coordinates {
		(1,8.18)
		(2,8.14)
		(5,8.06)
		(10,8.02)
		(20,7.94)
		(50,7.92)
		(100,7.91)		
	};
    \addlegendentry{VAE conversion} 
    
\end{axis} 
\end{tikzpicture}
\vspace{-4mm}
\caption{Effects of varying number of training sentences from 1 to 100\label{fig:objesent}}
\vspace{-5mm}
\end{figure}
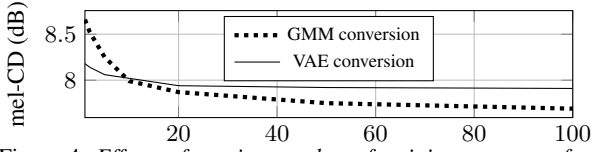

\vspace{-2mm}
We investigate the effect of VC training data size on the performance of the system. In order to be able to do objective test using mel-CD \cite{toda2007voice}, we require parallel data from the speakers. We use 20 parallel CMU-acric utterance from each speaker for computing the objective score. We vary non-parallel sentence numbers from source and target speaker that is used to compute the speaker embeddings. The results are shown in Figure~\ref{fig:objesent}. As can be seen, VAE performs better with less than 10 sentences, however, with more than 10 sentences, GMM starts achieving lower mel-CD. This might be due to VAE only having one degree of freedom (speaker identity vector) to convert the voice, compared to GMM which is able to use all the training data to adapt the background GMM model to better match the speaker data distribution.

\vspace{-4mm}

\subsection{Subjective evaluation}
\vspace{-2mm}
To subjectively evaluate voice conversion performance, we performed two perceptual tests. The first test measured speech quality, designed to answer the question ``how natural does the converted speech sound''?, and the second test measured speaker similarity, designed to answer the question ``how accurate does the converted speech mimic the target speaker"?. The listening experiments were carried out using Amazon Mechanical Turk, with participants who had approval ratings of at least 90\% and were located in North America. Both perceptual tests used three trivial-to-judge trials, added to the experiment to exclude unreliable listeners from statistical analysis. No listeners were flagged as unreliable in our experiments. In this subjective experiment, we focus on VAE train on TIMIT. We provide samples of VAE trained on TIMIT, CH, TIMIT+CH and VCTK in the samples webpage demo as well. In informal listening tests, we found out that VAE trained on TIMIT performs better than VAE trained on VCTK. Also VAE trained on TIMIT+CH generate better quality than on TIMIT or CH alone.
\vspace{-3mm}
\subsubsection{Speech quality}
\begin{figure}[t]
\centering
\begin{tikzpicture} 
\begin{axis}[
	bar width=0.42cm,
	width=7cm,
	height=3cm,
	ymax=1.0,
	ymin=0.2,
    ybar,
    grid=major,
    legend style={at={(0.5,1.25)},
      anchor=north,legend columns=-1},
    symbolic x coords={all,E2E,E2Z,Z2E, Z2Z},
    xtick=data,
    nodes near coords align={vertical},
    ylabel={MOS},
	ylabel near ticks,
    ] 
\addplot[fill=gray] coordinates                         {(all, 0.6125) (E2E,0.80625) (E2Z,0.6125) (Z2E,0.36875) (Z2Z,0.6625)}; 
\legend{GMMvsVAE} 
\end{axis} 
\end{tikzpicture}

\begin{tikzpicture} 
\begin{axis}[
	bar width=0.42cm,
	width=7cm,
	height=3cm,
	ymax=1.0,
	ymin=0.2,
    ybar,
    grid=major,
    legend style={at={(0.5,1.25)},
      anchor=north,legend columns=-1},
    symbolic x coords={all,M2F,F2F,F2M, M2M},
    xtick=data,
    nodes near coords align={vertical},
    ylabel={MOS},
	ylabel near ticks,
    ] 
\addplot[fill=gray] coordinates                         {(all, 0.6125) (F2F,0.975) (F2M,0.6125) (M2F,0.44375) (M2M,0.41875)};

\end{axis} 
\end{tikzpicture}

\caption{Speech Quality average score with gender and language break-down. Positive scores favor VAE. (confidence intervals for all is close to 0.13, and all scores are statistically significant)\label{fig:subjqual}}
\vspace{-4mm}
\end{figure}
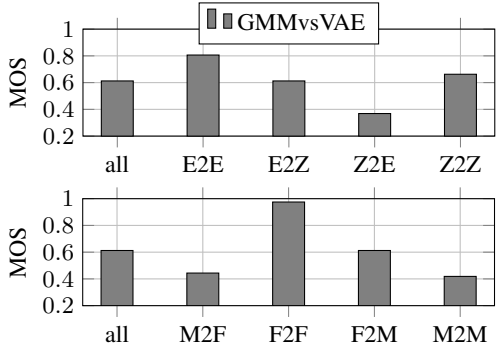

\vspace{-1.5mm}
To evaluate the speech quality of the converted utterances, we conducted a Comparative Mean Opinion Score (CMOS) test. In this test, listeners heard two stimuli A and B with the same content, generated using the same source speaker, but in two different processing conditions, and were then asked to indicate whether they thought B was better or worse than A, using a five-point scale comprised of +2 (much better), +1 (somewhat better), 0 (same), -1 (somewhat worse), -2 (much worse). We randomized the order of stimulus presentation, both the order of A and B, as well as the order of the comparison pairs. We utilized three processing conditions: GMM, VAE-STFT, VAE. We ran two separate experiments. First, to assess the effect of using World vocoder instead of STFT, we directly compared VAE-STFT vs. VAE. We only limited this experiment to English to English conversion. The experiment was administered to 40 listeners with each listener judging 16 sentence pairs. The results shows a very significant preference of VAE over VAE-STFT, achieving +1.25$\pm$0.12 mean score towards VAE. We performed planned one-sample t-tests with a mean of zero and achieved $p<0.0001$. Second, we assessed the VC approach effect by directly comparing GMM vs. VAE utterances. The experiment was administered to 40 listeners with each listener judging 80 sentence pairs. The results shows VAE has a statistically significant quality improvement over GMM, achieving +0.61$\pm$0.14 mean score towards VAE.  The language-breakdown of the results are shown in Figure~\ref{fig:subjqual}. We performed planned one-sample t-tests with a mean of zero and achieved $p<0.05$ for all language and gender conversion pairs separately, showing statistically significant improvements for all break-downs of gender and language. The Z2E conversion is achieving lowered quality compared to other conversion pairs. We speculate the reason is the slight noise present in THCHS-30 recordings which cause some distortion during vocoding.

\vspace{-3mm}
\subsubsection{Speaker similarity}
\begin{figure}[t]
\centering
\begin{tikzpicture} 
\begin{axis}[
	bar width=0.42cm,
	width=7cm,
	height=3.0cm,
	ymax=0.5,
	ymin=-0.7,
    ybar,
    grid=major,
    legend style={at={(0.5,1.25)},
      anchor=north,legend columns=-1},
    symbolic x coords={all,E2E,E2Z,Z2E, Z2Z},
    xtick=data,
    nodes near coords align={vertical},
    ylabel={MOS},
	ylabel near ticks,
    ] 
\addplot[fill=gray] coordinates                         {(all, -0.18427350427) (E2E,0.23384615385) (E2Z,-0.6358974359) (Z2E,-0.6358974359) (Z2Z,-0.017094017094)}; 
\addplot[fill=darkgray] coordinates                         {(all, -0.12243589743) (E2E,0.25641025641) (E2Z,-0.36153846154) (Z2E,-0.34333333333) (Z2Z,0.05128205128)}; 

\legend{GMM, VAE} 
\end{axis} 
\end{tikzpicture}

\vspace{-2.5mm}
\caption{Speech Similarity average score with language break-down. Positive scores are desirable. (Comparison of scores between GMM vs. VAE do not show statistical significance)\label{fig:subjsim}}
\vspace{-5mm}
\end{figure}
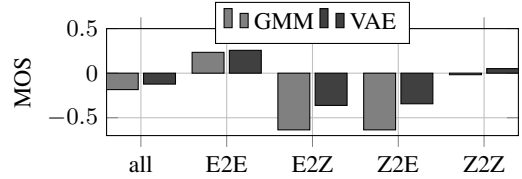

\vspace{-1.5mm}
To evaluate the speaker similarity of the converted utterances, we conducted a same-different speaker similarity test \cite{kain2001high}. In this test, listeners heard two stimuli A and B with different content, and were then asked to indicate whether they thought that A and B were spoken by the same, or by two different speakers, using a five-point scale comprised of +2 (definitely same), +1 (probably same), 0 (unsure), -1 (probably different), and -2 (definitely different). One of the stimuli in each pair was created by one of the two conversion methods, and the other stimulus was a purely MCEP-vocoded condition, used as the reference speaker. The listeners were explicitly instructed to disregard the language of the stimuli and merely judge based on the fact whether they think the utterances are from the same speaker regardless of the language. Half of all pairs were created with the reference speaker identical to the target speaker of the conversion (expecting listeners to reply ``same", ideally); the other half were created with the reference speaker being the same gender, but not identical to the target speaker of the conversion (expecting listeners to reply different). We only report ``same" scores. The experiment was administered to 40 listeners, with each listener judging 64 sentence pairs. The results are shown in Figure~\ref{fig:subjsim}. The results show GMM and VAE achieving  -0.18$\pm$0.15 and -0.12$\pm$0.16, respectively. We did not find any statistical significance between GMM vs VAE systems for average, or any of the language/gender conversion break-downs of the stimuli. For both VAE and GMM, we noticed that for E2E case, we achieve the highest average score and the only case that is able to transform identity, achieving $P<0.05$ in one-sample t-test compared to chance. This is reasonable given the training was done only on an English corpus. Furthermore, Z2Z achieves higher score compared to E2Z and Z2E. This might be due to the listener's bias toward not rating different language utterances as high score as same language utterances.

\vspace{-4mm}
\section{Conclusions}
\vspace{-1mm}
We proposed to exploit FHVAE model for challenging non-parallel and cross-lingual voice conversion, even with very small number of training utterances such as only one target speaker's utterance. We investigate the importance of speech representations and found that World vocoder outperformed STFT which was used in \cite{hsu2017unsupervised} in experimental evaluation, both speech quality and similarity. We also examined the effect of the size of training utterances from target speaker for VC, and our approach outperformed baseline with less than 10 sentences, and achieve reasonable performance even with only one training utterance. In the subjective tests, our approach achieved significantly better results than both VAE-STFT and GMM in speech quality, and outperformed VAE-STFT and comparable to GMM in speech similarity.   
As future work, we are interested in and working on joint end-to-end learning with Wavenet \cite{van2016wavenet} or Wavenet-vocoder \cite{tamamori2017speaker, hayashi2017investigation}, and also building models trained on multi-language corpora with or without explicit modeling of different languages such as providing language coding.


\bibliographystyle{IEEEtran}

\bibliography{reference}


\end{document}